\documentclass{PoS}

\newcommand{\agt}{\,\rlap{\lower 3.5 pt \hbox{$\mathchar \sim$}} \raise 1pt
 \hbox {$>$}\,}
\newcommand{\alt}{\,\rlap{\lower 3.5 pt \hbox{$\mathchar \sim$}} \raise 1pt
 \hbox {$<$}\,}

\boldmath
\title{Direct $J/\psi$ photoproduction at next-to-leading-order in
nonrelativistic QCD}
\unboldmath

\ShortTitle{Direct $J/\psi$ photoproduction in at NLO in NRQCD}

\author{Mathias Butensch\"on and \speaker{Bernd A. Kniehl}%
         \\
        II. Institut f\"ur Theoretische Physik, Universit\"at Hamburg,
        Luruper Chaussee 149, 22761 Hamburg\\
        E-mail: \email{mathias.butenschoen@desy.de},
                \email{kniehl@desy.de}
}

\abstract{
We calculate the cross section of inclusive direct $J/\psi$ photoproduction at
next-to-leading order within the factorization formalism of nonrelativistic
quantum chromodynamics, for the first time including the full relativistic
corrections due to the intermediate $^1\!S_0^{[8]}$, $^3\!S_1^{[8]}$, and
$^3\!P_J^{[8]}$ color-octet states.
A comparison of our results to recent H1 data suggests that the color octet
mechanism is indeed realized in $J/\psi$ photoproduction, although the
predictivity of our results still suffers from uncertainties in the
color-octet long-distance matrix elements.
}

\FullConference{XVIII International Workshop on Deep-Inelastic Scattering and Related Subjects\\
		 April 19--23, 2010\\
		 Convitto della Calza, Firenze, Italy}

\begin{document}

The factorization formalism of nonrelativistic quantum chromodynamics (NRQCD)
\cite{Bodwin:1994jh} provides a consistent theoretical framework for the
description of heavy-quarkonium production and decay.
This implies a separation of process-dependent short-distance coefficients, to
be calculated perturbatively as expansions in the strong-coupling constant
$\alpha_s$, from supposedly universal long-distance matrix elements
(LDMEs), to be extracted from experiment.
The relative importance of the latter can be estimated by means of velocity
scaling rules; {\it i.e.}, the LDMEs are predicted to scale with a definite
power of the heavy-quark ($Q$) velocity $v$ in the limit $v\ll1$.
In this way, the theoretical predictions are organized as double expansions in
$\alpha_s$ and $v$.
A crucial feature of this formalism is that it takes into account the complete
structure of the $Q\overline{Q}$ Fock space, which is spanned by the states
$n={}^{2S+1}L_J^{[a]}$ with definite spin $S$, orbital angular momentum
$L$, total angular momentum $J$, and color multiplicity $a=1,8$.
In particular, this formalism predicts the existence of color-octet (CO)
processes in nature.
This means that $Q\overline{Q}$ pairs are produced at short distances in
CO states and subsequently evolve into physical, color-singlet (CS) quarkonia
by the nonperturbative emission of soft gluons.
In the limit $v\to0$, the traditional CS model (CSM) is recovered in the case
of $S$-wave quarkonia.

Fifteen years after the introduction of the NRQCD factorization formalism
\cite{Bodwin:1994jh}, the existence of CO processes and the universality of the
LDMEs are still at issue and far from proven, despite an impressive series of
experimental and theoretical endeavors.
The greatest success of NRQCD was that it was able to explain the $J/\psi$
hadroproduction yield at the Fermilab Tevatron \cite{Cho:1995vh}, while the
CSM prediction lies orders of magnitudes below the data, even if the latter
is evaluated at next-to-leading order (NLO) or beyond \cite{Campbell:2007ws}.
Also in the case of $J/\psi$ photoproduction at DESY HERA, the CSM cross
section significantly falls short of the data, as demonstrated by a recent NLO
analysis \cite{Artoisenet:2009xh} using up-to-date input parameters and
standard scale choices, leaving room for CO contributions
\cite{Cacciari:1996dg}.
Similarly, the $J/\psi$ yields measured in electroproduction at HERA and in
two-photon collisions at CERN LEP2 were shown
\cite{Kniehl:2001tk,Klasen:2001cu} to favor the presence of CO processes.
As for $J/\psi$ polarization in hadroproduction, neither the
leading-order (LO) NRQCD
prediction \cite{Braaten:1999qk}, nor the NLO CSM one \cite{Campbell:2007ws}
leads to an adequate description of the Tevaton data.
The situation is quite similar for the polarization in photoproduction at HERA
\cite{Artoisenet:2009xh}.

In order to convincingly establish the CO mechanism and the LDME universality,
it is an urgent task to complete the NLO description of $J/\psi$ hadro-
\cite{Campbell:2007ws} and photoproduction
\cite{Artoisenet:2009xh,Kramer:1994zi}, regarding both $J/\psi$ yield and
polarization, by including the full CO contributions at NLO.
While the NLO contributions due to the $^1\!S_0^{[8]}$ and $^3\!S_1^{[8]}$
CO states may be obtained using standard techniques \cite{Kramer:1994zi}, the
NLO treatment of $^3\!P_J^{[8]}$ states in $2\to2$ processes requires a more
advanced technology, which has been lacking so far.
In fact, the $^3\!P_J^{[8]}$ contributions represent the missing links in all
those previous NLO analyses
\cite{Campbell:2007ws,Artoisenet:2009xh,Kramer:1994zi},
and there is no reason at all to expect them to be insignificant.
Specifically, their calculation is far more intricate because the application
of the $^3\!P_J^{[8]}$ projection operators to the short-distance scattering
amplitudes produce particularly lengthy expressions involving complicated
tensor loop integrals and exhibiting an entangled pattern of infrared 
singularities.
This technical bottleneck, which has prevented essential progress in the global
test of NRQCD factorization for the past fifteen years, was overcome for
the first time in Ref.~\cite{Butenschoen:2009zy}, which we review here.

\begin{figure}
\includegraphics[width=6cm]{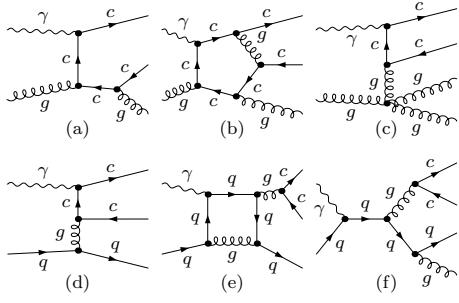}
\caption{Sample diagrams contributing at LO (a and d) and
to the virtual (b and e) and real (c and f) NLO corrections.}
\label{fig:Examples} 
\end{figure}

In direct photoproduction, a quasi-real photon $\gamma$ that is radiated off
the incoming electron $e$ interacts with a parton $i$ stemming from the
incoming proton $p$.
Invoking the Weizs\"acker-Williams approximation and the factorization theorems
of the QCD parton model and NRQCD \cite{Bodwin:1994jh}, the inclusive $J/\psi$
photoproduction cross section is evaluated from
\begin{equation}
d\sigma(ep\to J/\psi+X)
=\sum_{i,n} \int dxdy\, f_{\gamma/e}(x)f_{i/p}(y)
\langle{\cal O}^{J/\psi}[n]\rangle
d\sigma(\gamma i\to c\overline{c}[n]+X),
\label{Overview.Cross}
\end{equation}
where $f_{\gamma/e}(x)$ is the photon flux function, $f_{i/p}(y)$ are the
parton distribution functions (PDFs) of the proton,
$\langle{\cal O}^{J/\psi}[n]\rangle$ are the LDMEs, and
$d\sigma(\gamma i\to c\overline{c}[n]+X)$ are the partonic cross sections.
Working in the fixed-flavor-number scheme, $i$ runs over the gluon $g$ and the
light quarks $q=u,d,s$ and anti-quarks $\overline q$.
The Fock states contributing through the order of our calculation include
$n={}^3S_1^{[1]},{}^1S_0^{[8]},{}^3S_1^{[8]},{}^3P_J^{[8]}$.
Example Feynman diagrams for partonic LO subprocesses 
$\gamma i\to c\overline{c}[n]+X$ as well as virtual- and
real-correction diagrams are shown in Fig.~\ref{fig:Examples}.

We now describe our theoretical input and the kinematic conditions for our
numerical analysis.
We set $m_c=m_{J/\psi}/2$, adopt the values of $m_{J/\psi}$, $m_e$, and
$\alpha$ from Ref.~\cite{Amsler:2008zzb}, and use the one-loop (two-loop)
formula for $\alpha_s^{(n_f)}(\mu)$, with $n_f=3$ active quark flavors, at LO
(NLO).
As for the proton PDFs, we use set CTEQ6L1 (CTEQ6M) \cite{Pumplin:2002vw} at LO
(NLO), which comes with an asymptotic scale parameter of
$\Lambda_\mathrm{QCD}^{(4)}=215$~MeV (326~MeV), so that
$\Lambda_\mathrm{QCD}^{(3)}=249$~MeV (389~MeV).
We evaluate the photon flux function using Eq.~(5) of Ref.~\cite{Kniehl:1996we}
with the cut-off $Q_\mathrm{max}^2=2$~GeV$^2$ \cite{Adloff:2002ex,H1.prelim} on
the photon virtuality.
Our default choices for the renormalization, factorization, and NRQCD scales
are $\mu_r=\mu_f=m_T$ and $\mu_\Lambda=m_c$, respectively, where
$m_T=\sqrt{p_T^2+4m_c^2}$ is the $J/\psi$ transverse mass.
We adopt the LDMEs from Ref.~\cite{Kniehl:1998qy}, which were fitted to
Tevatron~I data using the CTEQ4 PDFs, because, besides the usual LO set, they
also comprise a {\em higher-order-improved} set determined by approximately
taking into account dominant higher-order effects due to multiple-gluon
radiation in inclusive $J/\psi$ hadroproduction, which had been found to be
substantial by a Monte Carlo study \cite{CanoColoma:1997rn}.
We disentangle $\langle {\cal O}^{J/\psi}(^1\!S_0^{[8]}) \rangle$ and
$\langle {\cal O}^{J/\psi}(^3\!P_0^{[8]}) \rangle$, a linear combination of
which is fixed by the fit only, as in Ref.~\cite{Klasen:2004tz}.
The LO CO LDMEs are similar to the those obtained in Ref.~\cite{Kniehl:2006qq}
by fitting Tevatron~II data using the CTEQ6L1 PDFs \cite{Pumplin:2002vw}.
The higher-order-improved CO LDMEs are likely to undershoot the genuine ones,
which are presently unknown.

Recently, the H1 Collaboration presented preliminary data on inclusive $J/\psi$
photoproduction taken in collisions of 27.6~GeV electrons or positrons on
920~GeV protons in the HERA~II laboratory frame \cite{H1.prelim}.
They nicely agree with their previous measurement at HERA~I
\cite{Adloff:2002ex}.
These data come as singly differential cross sections in $p_T^2$,  
$W=\sqrt{(p_\gamma+p_p)^2}$, and $z=(p_{J/\psi}\cdot p_p)/(p_\gamma\cdot p_p)$,
in each case with certain acceptance cuts on the other two variables.
Here, $p_\gamma$, $p_p$, and $p_{J/\psi}$ are the photon, proton, and $J/\psi$
four-momenta, respectively.
In the comparisons below, we impose the same kinematic conditions on our
theoretical predictions.

\begin{figure*}
\begin{tabular}{ccc}
\includegraphics[width=4.7cm]{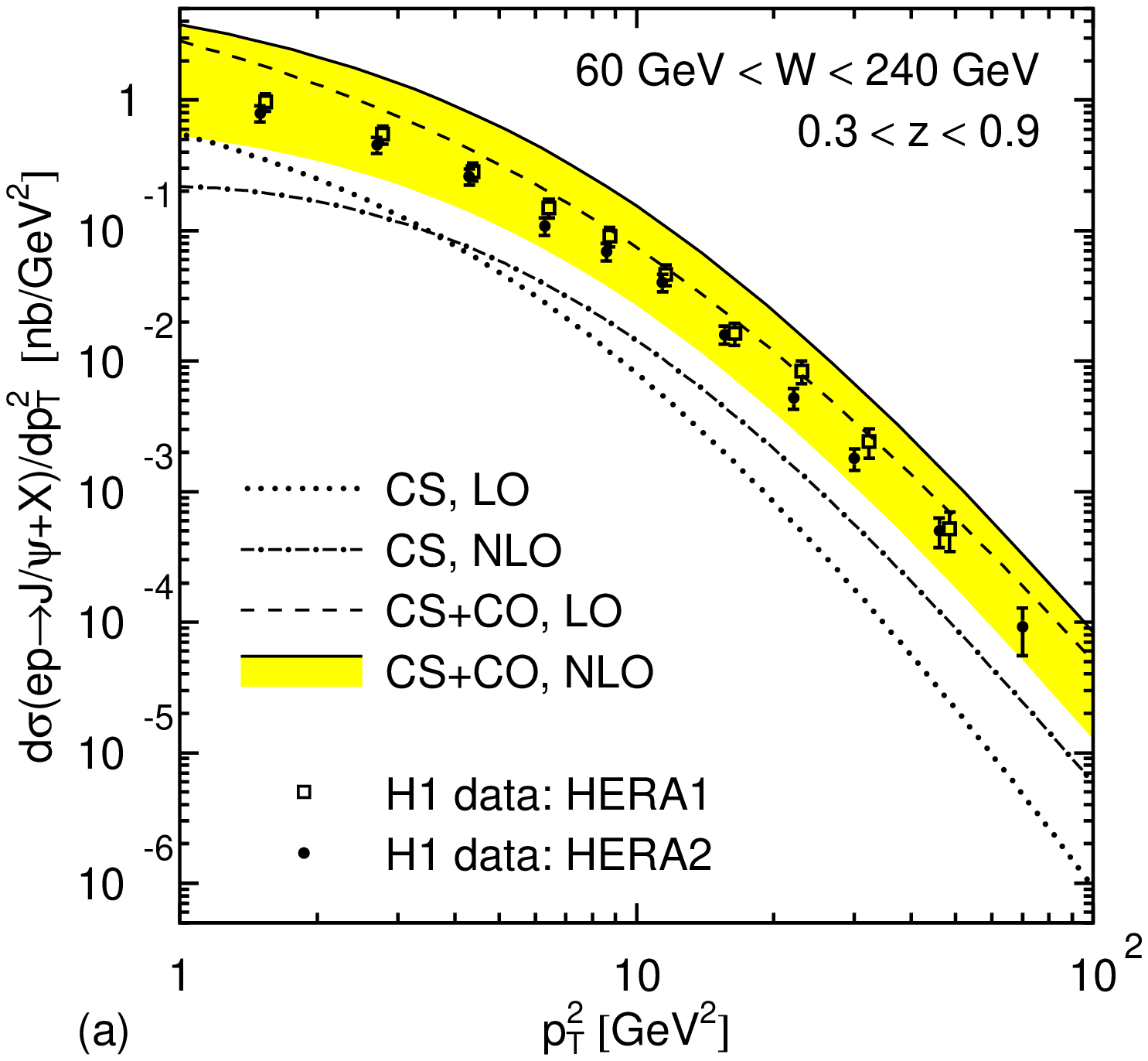}
&
\includegraphics[width=4.7cm]{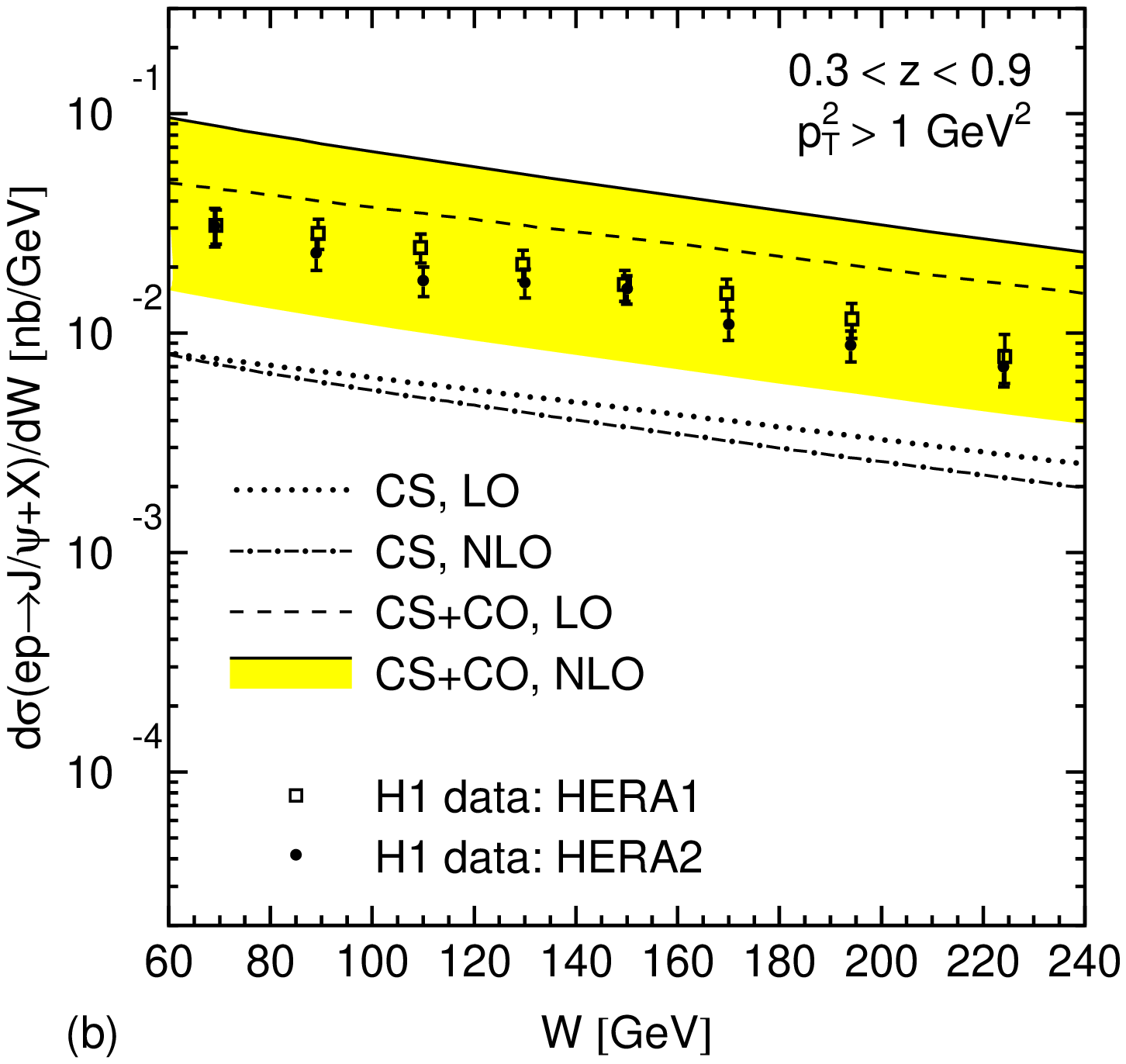}
&
\includegraphics[width=4.7cm]{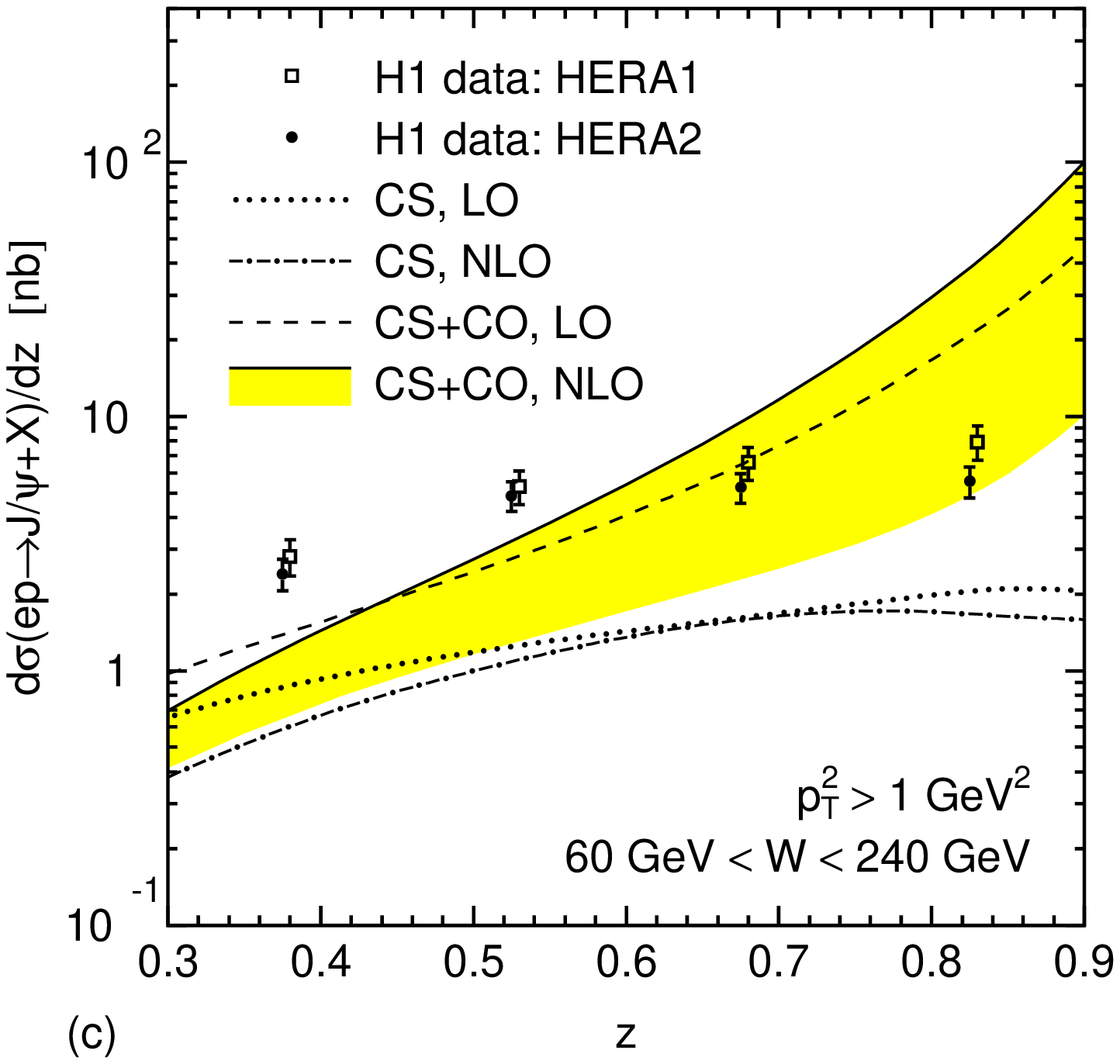}
\end{tabular}
\caption{(a) $p_T^2$, (b) $W$, and (c) $z$ distributions of inclusive $J/\psi$
photoproduction at LO and NLO in the CSM and full NRQCD in comparison with H1
data \cite{Adloff:2002ex,H1.prelim}.
The shaded (yellow) bands indicate the theoretical uncertainty due to the CO
LDMEs.}
\label{fig:results}
\end{figure*}

The H1 measurements \cite{Adloff:2002ex,H1.prelim} of the $p_T^2$, $W$, and $z$
distributions of inclusive $J/\psi$ photoproduction are compared with our new
NLO predictions in full NRQCD in Fig.~\ref{fig:results}(a)--(c), respectively. 
The uncertainty due the LDMEs is indicated by shaded (yellow) bands, whose
upper margins (solid lines) refer to the LO set.
For comparison, also the default predictions at LO (dashed lines) as well as
those of the CSM at NLO (dot-dashed lines) and LO (dotted lines) are shown.
Notice that the experimental data are contaminated by the feed-down from
heavier charmonia, mainly due to $\psi^\prime\to J/\psi+X$, which yields an
estimated enhancement by about 15\% \cite{Kramer:1994zi}.
Furthermore, our predictions do not include resolved photoproduction, which
contributes appreciably only at $z\alt0.3$ \cite{Kniehl:1998qy}, and
diffractive production, which is confined to the quasi-elastic domain at
$z\approx1$ and $p_T\approx0$.
These contributions are efficiently suppressed by the cut $0.3<z<0.9$ in
Figs.~\ref{fig:results}(a) and (b), so that our comparisons are indeed
meaningful.
We observe that the NLO corrections enhance the NRQCD cross section, by up to
115\%, in the kinematic range considered, except for $z\alt0.45$, where they
are negative.
As may be seen from Fig.~\ref{fig:results}(c), the familiar growth of the LO
NRQCD prediction in the upper endpoint region, leading to a breakdown at
$z=1$, is further enhanced at NLO.
The solution to this problem clearly lies beyond the fixed-order treatment and
may be found in soft collinear effective theory \cite{Fleming:2006cd}.
The experimental data are nicely gathered in the central region of the error
bands, except for the two low-$z$ points in Fig.~\ref{fig:results}(c), which
overshoot the NLO NRQCD prediction.
However, this apparent disagreement is expected to fade away once the
NLO-corrected NRQCD contribution due to resolved photoproduction is included.
In fact, the above considerations concerning the large size of the NLO
corrections to hadroproduction directly carry over to resolved
photoproduction, which proceeds through the same partonic subprocesses.
On the other hand, the default CSM predictions significantly undershoot the
experimental data, by typically a factor of 4, which has already been observed
in Ref.~\cite{Artoisenet:2009xh}.
Except for $p_T^2\agt4$~GeV$^2$, the situation is even deteriorated by the
inclusion of the NLO corrections.

Despite the caveat concerning our limited knowledge of the CO LDMEs at NLO,
we conclude that the H1 data \cite{Adloff:2002ex,H1.prelim} show clear evidence
of the existence of CO processes in nature, as predicted by NRQCD, supporting
the conclusions previously reached for hadroproduction at the Tevatron
\cite{Cho:1995vh} and two-photon collisions at LEP2 \cite{Klasen:2001cu}.
In order to further substantiate this argument, it is indispensable to
complete the NLO analysis of inclusive $J/\psi$ hadroproduction in NRQCD, by
treating also the $^3\!P_J^{[8]}$ channels at NLO, so as to permit a genuine
NLO fit of the relevant CO LDMEs to Tevatron and CERN LHC data.
This goal is greatly facilitated by the technical advancement achieved in the
present analysis.

\acknowledgments

This work was supported in part by BMBF Grant No.\ 05H09GUE, DFG Grant
No.\ KN~365/6--1, and HGF Grant No.\ HA~101.

\end{document}